\documentclass[twoside,slac_one]{revtex4}
\usepackage{graphicx}
\usepackage{fancyhdr}
\usepackage{amsmath} 
\usepackage{bm}
\usepackage{amsxtra}
\usepackage{amssymb}
\usepackage{amsthm}
\usepackage{latexsym}
\usepackage{lscape}

\begin{document}

\title{Review of the measurements of the anomalous like-sign dimuon charge asymmetry
in $p \bar{p}$ collisions by the D\O\ Collaboration}

%

\author{B. Hoeneisen}
\affiliation{Universidad San Francisco de Quito, Ecuador}

\begin{abstract}
In this short review we present the history, 
an overview the analysis, and some personal comments on the
anomalous like-sign dimuon charge asymmetry measurements
by the D\O\ Collaboration.
\end{abstract}

\maketitle

\thispagestyle{fancy}


\section{\label{Introduction} Introduction}
The D\O\ Collaboration has published three measurements of the
like-sign dimuon charge asymmetry in proton-antiproton collisions:
in 2006 with 1.0 fb$^{-1}$ of data \cite{D01}, in
2010 with 6.1 fb$^{-1}$ \cite{D02}, and in 2011 with 9.0 fb$^{-1}$ \cite{D03}.
The results differ from the Standard Model prediction by 1.7, 3.2
and 3.9 standard deviations respectively, so have attracted much 
attention. Is this, at long last, the first hint of new physics
beyond the Standard Model of quarks and leptons?
Are we seeing a new source of CP violation that could explain
why the Universe has matter? If so, is this asymmetry due to $B^0_d \bar{B}^0_d$
mixing, $B^0_s	\bar{B}^0_s$ mixing, both or none? The experimental
situation is changing rapidly. The Tevatron has been turned off for the
last time and we are now doing the final analysis with the full data set
of about 10.0 fb$^{-1}$. On the other side of the pond, the LHC is collecting
data at a rate exceeding all estimates of the preceding year. The LHCb 
Collaboration is showing results on B-physics with unprecedented low
backgrounds and high precision, and the results on 
$B^0_s (\bar{B}^0_s) \rightarrow J/\psi \phi$ presented at the Lepton Photon
conference in September of 2011 have been unexpected.

In this short review we present the history, an overview of the analysis, 
and some personal comments on the like-sign dimuon charge asymmetry. 
We do not enter into any of the technical details, because they would fill a
book and are readily available: full descriptions of the measurements can be found in
\cite{D01, D02, D03}, the Proceedings of the DPF-2011 Conference are in \cite{BH},
and the theory of CP-, T- and CPT-violation in the
mixing, decay, and interference of mixing and decay of $B^0_q$ ($q = d$ or $s$) mesons
can be found in \cite{CPV}.

\section{\label{History} History}

At the Fermilab Tevatron collider, $b$ quarks are produced mainly in $b \bar{b}$ pairs.
Therefore, to observe an event with two like-sign muons
from semi-leptonic $b$-hadron decay, one
of the hadrons must be a $B^0_d$ or $B^0_s$ meson that oscillates and decays to
a muon of charge opposite of that of the original $b$ quark.
The oscillation $B^0_q \leftrightarrow \bar{B}^0_q$ is described by
``box" Feynman diagrams. To calculate these diagrams it is necessary to 
integrate over the 4-momentum running around the loop. These integrals can pick up
new virtual particles of high mass not directly accessible at the Tevatron.
New particles add new Feynman box diagrams, changing the Standard Model
matrix element $M^{q,\textrm{SM}}_{12}$, that induces  $B^0_q \bar{B}^0_q$ mixing, 
to $M^{q,\textrm{SM}}_{12} |\Delta_q| e^{i \phi^\Delta_q}$.
A non-zero phase $\phi^{\Delta}_q$ would be a new source of CP violation.
For example, a 4th generation of quarks
and leptons would contribute new box diagrams with a $t'$ in the loop.
The CKM matrix would become $4 \times 4$, and have 3 observable CP violating phases,
compared with 1 in the $3 \times 3$ matrix of the Standard Model.
A second example is a new, perhaps right handed, $SU_2$ symmetry, with a
new CKM matrix and a new CP violating phase, with the
$W'$ contributing new box diagrams. A third example are charged higgs bosons
in the loop.

The like-sign dimuon charge asymmetry from semi-leptonic decay of $b$-hadrons,
\begin{equation}
A^b_\textrm{sl} \equiv \frac{N^{++}_b - N^{--}_b}{N^{++}_b + N^{--}_b},
\label{A}
\end{equation}
has contributions from the semi-leptonic charge asymmetries
$a^d_\textrm{sl}$ and $a^s_\textrm{sl}$ of $B^0_d$ and $B^0_s$
mesons:
\begin{eqnarray}
A^b_\textrm{sl} & = & C_d a^d_\textrm{sl} + C_s a^s_\textrm{sl}.
\label{Ab}
\end{eqnarray}

Our first attempt to measure the like-sign dimuon charge asymmetry in Run I
of the D\O\ detector (1992 to 1996) did 
not succeed mainly because of the background from the
Main Ring accelerator that went right through our calorimeter!
The D\O\ detector was shut down and upgraded from 1996 to 2002.
The Main Ring was removed; a superconducting solenoid
was installed around the collision point; the tracking wire chambers were 
replaced by a silicon micro vertex detector and a scintillating fiber tracker;
the forward muon system was replaced by mini-drift chambers; and the
D\O\ detector was fully covered by scintillating trigger counters that 
reduced the cosmic ray and halo backgrounds.

The upgraded D\O\ detector was uniquely well suited for the precision measurement
of the like-sign dimuon charge asymmetry
for the following reasons: (i) the initial $p \bar{p}$ state
was symmetric with respect to CP conjugation;
(ii) the solenoid and toroid magnetic fields were reversed
periodically, thereby canceling first order detector
asymmetries; (iii) the shielding between the central tracker
and the outer muon spectrometer was sufficient to reduce the
background from hadron punch-through to the 1\% level; and (iv)
the muon track parameters were measured twice, once by the central
tracker, and once by the outer muon system, thereby
reducing fake muon tracks and kaon and pion decay backgrounds, and allowing
the detector asymmetries and backgrounds to be studied in more detail.
 
The pioneering measurement published in 2006 \cite{D01} had to establish which
issues were important, and which could be neglected.
How should we parametrize the detector charge asymmetries? How could we measure
these parameters with data? Do positive and negative muons have the same energy loss
in the calorimeter and iron toroids to one part in $10^4$?
Is the efficiency of detecting, triggering and tracking positive and negative
muons the same? 
What is the fraction and charge asymmetry of fake muon tracks, of cosmic rays
detected once, of cosmic rays detected twice (once entering the 
D\O\ detector, and once exiting it), and of the kaon decay background?
What is the fraction of tracks that have a  wrong charge measurement? 
Is the toroid magnetic
field the same for both polarities to one part in $10^{4}$?
Is there any charge bias in the track reconstruction software?
Is muon production forward-backward symmetric?
And on and on. 
Submitting Monte Carlo jobs daily for six months
resulted in only 21 dimuon events, which was barely sufficient for this analysis. 

The measurement of 2010 was greatly improved. The kaon, pion and punch-trough
backgrounds were measured by reconstructing exclusive decays in the
same inclusive muon and like-sign dimuon data sets, with minimal use of Monte
Carlo simulation. The residual muon spectrometer
charge asymmetry was measured by reconstructing $J/\psi$'s using only the central
detector tracks. The residual charge asymmetry of the central tracker
was measured by counting positive and negative particles and correcting
for kaon decay. All of these measurements were done as a function of
the momentum $p_T$ of the particles transverse to the proton-antiproton beams.
The main cross-check was the measurement of the charge asymmetry
of inclusive muons. There were about 300 inclusive muons
per like-sign dimuon event. The charge asymmetry of the inclusive muons
was dominated by the residual detector asymmetry (after averaging over
the 4 solenoid-toroid magnet polarities) and by the kaon decay background,
because in this data set any asymmetry from $B^0_q \bar{B}^0_q$ mixing
was diluted by decays without mixing. The inclusive muon charge
asymmetry provided an indispensable ``closure test".
This closure test was also done separately for each bin of $p_T$. These improvements were
necessary to keep the systematic uncertainty below the statistical one.

Let us now describe in more detail the measurement of 2011.

\section{\label{Overview} Overview of the measurement with 9 fb$^{-1}$}
Two data sets were used for this measurement:
the \textit{inclusive muon set}, collected with single muon triggers,
had $n^+ + n^- = 2.04 \times 10^9$ muon candidates passing
strict quality selections; and the \textit{like-sign dimuon set},
collected with dimuon triggers, had
$N^{++} + N^{--} = 6.02 \times 10^6$ 
dimuon events with each muon passing the
same quality selections, and in addition the following dimuon
requirements: same charge sign, same associated vertex, and
a dimuon invariant mass greater than 2.8 GeV to suppress
events with the two muons coming from the same B-hadron decay
cascade. Counting inclusive muons, and like-sign dimuons, we
obtained the ``raw" asymmetries
\begin{eqnarray}
a & \equiv & \frac{n^+ - n^-}{n^+ + n^-} = (+0.688 \pm 0.002)\%, \textrm{ and} \label{a1} \\
A & \equiv & \frac{N^{++} - N^{--}}{N^{++} + N^{--}} = (+0.126 \pm 0.041)\%.
\end{eqnarray}
These ``raw" asymmetries were corrected for kaon, pion and proton decay or
punch-through, and for the residual muon detector asymmetry after averaging
over the 4 solenoid-toroid magnet polarity combinations. These corrections
were measured, as a function of $p_T$, with the same data sets, by reconstructing
exclusive decays, with minimal use of simulation.
As far as possible, measurements were redundant using two independent channels.
The main background asymmetry was due to kaon decay. Positive
kaons had a longer inelastic interaction length in the calorimeter
than negative kaons, and hence had more time to decay.
The resulting positive charge asymmetry contributions from kaon decay were measured to
be $(+0.776 \pm 0.021)\%$ for $a$, and $(+0.633 \pm 0.031)\%$ for $A$.
The residual muon detector asymmetries were measured (reconstructing $J/\psi$'s
from central detector tracks) to be
$(-0.047 \pm 0.012)\%$ for $a$, and $(-0.212 \pm 0.030)\%$ for
$A$. Corrections due to pion decay and proton punch-through were
smaller. The charge asymmetries, corrected for background and detector
effects, are
\begin{eqnarray}
a - a_{\textrm{bkg}} & \equiv & (-0.034 \pm 0.042\textrm{ (stat)})\%, \textrm{ and}\label{a} \\
A - A_{\textrm{bkg}} & \equiv & (-0.276 \pm 0.067\textrm{ (stat)})\%. \label{b}
\end{eqnarray}

\begin{figure}[ht]
\centering
\includegraphics[width=80mm]{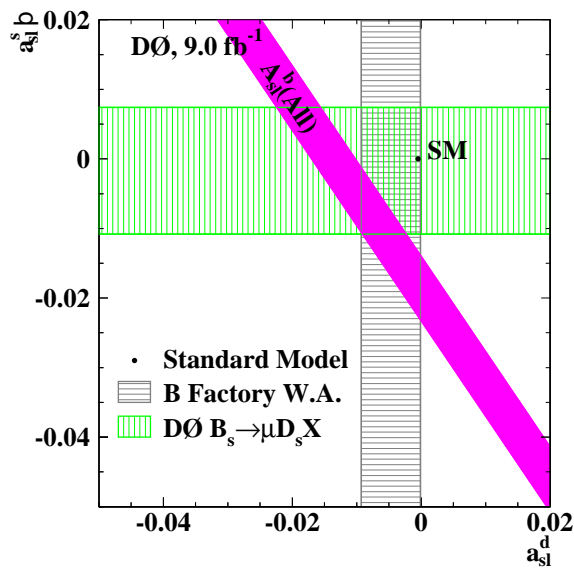}
\caption{Comparison of $A^b_\textrm{sl}$ in data with the
Standard Model prediction for $a^d_\textrm{sl}$ and $a^s_\textrm{sl}$.
Also shown are the measurements of $a^d_\textrm{sl}$~\cite{hfag} and
$a^s_\textrm{sl}$~\cite{asl-d0}. The bands represent the $\pm 1$
standard deviation uncertainties on
each individual measurement. } 
\label{1}
\end{figure}

We interpreted these charge asymmetries as arising from CP
violation in the mixing of $B^0_d$ and $B^0_s$ mesons.
To obtain $A^b_\textrm{sl}$, we divided the corrected asymmetries 
$a - a_{\textrm{bkg}}$ and $A - A_{\textrm{bkg}}$ by ``dilution
factors"
\begin{eqnarray}
c_b & = & +0.061 \pm 0.007, \textrm{ and} \\
C_b & = & +0.474 \pm 0.032,
\end{eqnarray}
respectively. These ``dilutions factors", obtained from 
simulation, are due to prompt decays that are
not direct semi-leptonic $b \rightarrow \mu X$, i.e.
sequential decays $b \rightarrow c \rightarrow \mu X$,
decays with $b \rightarrow c \bar{c} q$ with $c \rightarrow \mu X$
or $\bar{c} \rightarrow \mu X$, decays of light mesons,
events with $c \bar{c}$, and events with $b \bar{b} c \bar{c}$.
The results are
\begin{eqnarray}
A^b_\textrm{sl} & = & (-1.04 \pm 1.30~({\rm stat}) \pm 2.31~({\rm syst})) \%, \textrm{ and} \\
A^b_\textrm{sl}	& = & (-0.808 \pm 0.202~({\rm stat}) \pm 0.222~({\rm syst})) \%,
\end{eqnarray}
respectively. The asymmetries $a$ and $A$ have correlated backgrounds.
Therefore, a more precise measurement of $A^b_\textrm{sl}$ can be obtained from
$A - \alpha a$. The parameter $\alpha = 0.89$ was chosen to minimize the total 
uncertainty of $A^b_\textrm{sl}$. The resulting final measurement is
\begin{eqnarray}
A^b_\textrm{sl}	& = & (-0.787 \pm 0.172~({\rm stat}) \pm 0.093~({\rm syst}))\%.
\label{Asl_final}
\end{eqnarray}
This result differs from the Standard Model prediction \cite{D03}, 
\begin{equation}
A^b_\textrm{sl} = (-0.028^{+0.005}_{-0.006}) \%, 
\end{equation}
by 3.9 standard deviations. Equation (\ref{Ab}) with the result (\ref{Asl_final})
is show in Figure \ref{1}.

The ``residual" like-sign dimuon charge asymmetry, obtained from $A - \alpha a$, 
corrected for detector and background
effects, but without any interpretation, i.e. without dividing by a dilution
factor, is
\begin{equation}
A_\textrm{res} = (-0.246 \pm 0.052~({\rm stat}) \pm 0.021~({\rm syst}))\%,
\label{Ares}
\end{equation}
and differs from the Standard Model prediction by 4.2 standard deviations.

These results are in good agreement with the previous measurements:
the publication of 2010 with 6.1 fb$^{-1}$ reported a like-sign dimuon
charge asymmetry from semi-leptonic decay \cite{D02}
\begin{eqnarray}
A^b_\textrm{sl} & = & (-0.957 \pm 0.251~({\rm stat}) \pm 0.146~({\rm syst}))\%;
\label{Asl_2010}
\end{eqnarray}
and the measurement of 2006 with 1.0 fb$^{-1}$ \cite{D01} reported a
``residual" charge asymmetry, corrected for detector and background effects,
\begin{equation}
A_\textrm{res} = (-0.28 \pm 0.13~({\rm stat}) \pm 0.09~({\rm syst}))\%.
\label{Ares_2006}
\end{equation}

\begin{figure}[ht]
\centering
\includegraphics[width=0.50\textwidth]{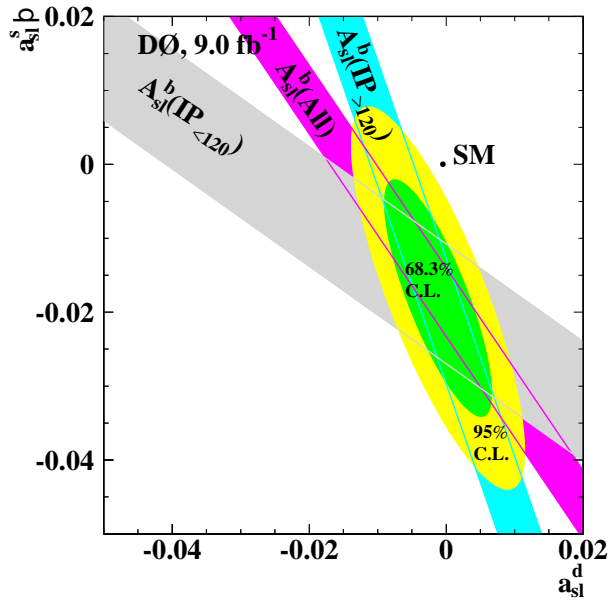}
\caption{Measurements of $A^b_\textrm{sl}$ with different muon impact parameter
selections in the $(a^d_\textrm{sl}, a^s_\textrm{sl})$ plane.
The bands represent the $\pm 1$ standard deviation uncertainties on each
individual measurement. The ellipses represents the 
$68.3\%$ and $95\%$ confidence-level contours of
$a^d_\textrm{sl}$ and $a^s_\textrm{sl}$ values obtained from the measurements
with IP selections. 
}
\label{2}
\end{figure}

To explore the origin of the charge asymmetry,
we performed measurements with muon impact parameter IP $> 120 \mu$m and
IP $< 120 \mu$m (for like-sign dimuons each muon is required to pass
the IP cut). IP is the distance of closest approach of the muon track
to the primary vertex projected onto
the plane transverse to the $p \bar{p}$ beams.
The coefficients $C_d$ and $C_s$ in (\ref{Ab})
depend on the IP cut, since for IP $> 120 \mu$m the $B^0_d$-meson
has a longer lifetime on average, and hence a greater probability
to oscillate. The results of these measurements are consistent with the
hypothesis of CP violation in the mixing of $B^0_d$ and $B^0_s$
mesons with semi-leptonic decay asymmetries
\begin{eqnarray}
a^d_\textrm{sl} & = & (-0.12 \pm 0.51) \%, \textrm{ and} \\
a^s_\textrm{sl} & = & (-1.81 \pm 1.04) \%.
\end{eqnarray}
These two asymmetries are correlated as shown in 
Figure \ref{2}.

\section{\label{Comments} Comments}
As data was collected over the years, more and more statistically significant
cross-checks became possible. Each of these tests could have revealed inconsistencies,
but none have been observed so far.

The ``raw" inclusive muon charge
asymmetry $a$ is dominated by background and detector asymmetries due to
the small value of the dilution factor $c_b$. Therefore, the corrected 
inclusive muon charge asymmetry (\ref{a})
serves as a ``closure test" of the measurements of the backgrounds
and detector asymmetries. Equation (\ref{a}) indicates that the sum of
uncertainties of all background and detector asymmetries (those that have
been explicitly considered, and even those that have not been imagined!)
is less than approximately $\pm 0.042 \%$, which is
smaller than the statistical uncertainty of $A - A_{\textrm{bkg}}$ in (\ref{b}).
The ``closure test" has been presented in \cite{D03} as a function
of transverse momentum $p_T$, and pseudo-rapidity $\eta$, and
good agreement is found.

Since background muons are mainly produced by decays of kaons 
and pions, their track parameters independently measured by the central
tracker and by the outer muon spectrometer can differ.
The background fractions therefore depend strongly on the
$\chi^2$ of the difference between these two measurements.
In test C of Table XV of \cite{D03} the $\chi^2$ cut is changed from
12 to 4 (for 4 degrees of freedom). The ``raw" charge asymmetry
$a$ ($A$) changes from +0.688\% to +0.325\% (+0.126\% to -0.361\%),
yet the measured value of $A^b_\textrm{sl}$ does not change
significantly (see \cite{D03} for details). Note that
$A$ changes sign with this reduction of background.

In Tables XV and XVI of \cite{D03} are presented 18
tests by varying the muon $p_T$, $\eta$ and $\phi$ ranges,
the muon quality selections, the triggers, the maximum impact
parameter, the instantaneous luminosity, using only
one pair of solenoid-toroid magnet polarities, and different
data running periods. 
The $\chi^2$ of these $18+1$ measurements of
$A^b_\textrm{sl}$, taking account
of common events, is 17.1 for 18 degrees of freedom.
These tests prove that the total uncertainty of
$A^b_\textrm{sl}$ is correct.

In cross-check K of \cite{D03} we repeat the measurement of
$A^b_{\textrm{sl}}$ using only central muons with pseudo-rapidity
$| \eta | < 1.6$. The result does not change significantly.
This cross-check is important because the central and forward
regions have independent detectors: the central muon detector uses
large proportional drift chambers, while the forward muon detector
uses mini drift tubes. These two muon systems have independent 
reconstruction software. The tracking system has silicon 
barrels and a scintillating fiber tracker
in the central region, while silicon discs are used in the forward
region. Even the magnetized iron toroids are different in the
central and forward regions.

Applying the impact parameter cut IP $> 120 \mu$m 
reduces the kaon and pion decay backgrounds by the large 
factors 3 to 5, the ``raw" charge asymmetry  
$a$ ($A$) changes from +0.688\%	to -0.014\% (+0.126\% to -0.529\%),
yet the results are again consistent.
Note that both asymmetries $a$ and $A$ change sign with
this reduction of the backgrounds, and the measured
$A^b_\textrm{sl}$ from $a$ 
($A^b_\textrm{sl} = (-0.422 \pm 0.240 \textrm{ (stat)} \pm 0.121 \textrm{ (syst)}) \%$) 
and $A$ 
($A^b_\textrm{sl} = (-0.818 \pm 0.342 \textrm{ (stat)} \pm 0.067 \textrm{ (syst)}) \%$) 
are compatible in spite of the very different dilution factors $c_b$ and $C_b$.

It is a challenge to imagine a background or detector effect that
can make both corrected charge asymmetries (\ref{a}) and (\ref{b}), which
are so different, zero simultaneously.
Finally, we note in Eq. (\ref{Asl_final}) that the uncertainty of
$A^b_\textrm{sl}$ is still dominated by statistics.

\section{\label{Anomalies} Anomalies in $B^0_d$ and $B^0_s$ mixing and decay}
New physics in the mixing of $B^0_q$ mesons, assuming CPT invariance, can be parametrized by
4 complex numbers $\Delta_q$ and $\tilde{\Delta}_q$
as follows \cite{Lenz}:
\begin{eqnarray}
M^q_{12} & \equiv & M^{q, \textrm{SM}}_{12} \cdot \Delta_q
= M^{q, \textrm{SM}}_{12} \cdot \left| \Delta_q \right| e^{i \phi^\Delta_q}, \textrm{ and } \\
\Gamma^q_{12} & \equiv & \Gamma^{q, \textrm{SM}}_{12} \cdot \tilde{\Delta}_q
= \Gamma^{q, \textrm{SM}}_{12} \cdot \left| \tilde{\Delta}_q \right| e^{-i \tilde{\phi}^\Delta_q},
\end{eqnarray}
with $q = d, s$.
In the presence of new physics, the semi-leptonic charge asymmetries, 
and the differences in mass and decay rates of the eigenstates are
\begin{eqnarray}
a^q_\textrm{sl} & = & \left| \frac{\Gamma^{q, SM}_{12}}{M^{q, SM}_{12}} 
\right| \cdot \frac{|\tilde{\Delta}_q|}{|\Delta_q|}
\sin{(\phi^\textrm{SM}_q + \phi^\Delta_q + \tilde{\phi}^\Delta_q)}
(1 + 2 \delta^q_A), \\
\Delta m_q & = & 2 | M^{q, SM}_{12} | \cdot | \Delta_q |, \textrm{ and } \\
\Delta \Gamma_q & = & 2 | \Gamma^{q, SM}_{12} | \cdot | \tilde{\Delta}_q | 
\cos (\phi^\textrm{SM}_q + \phi^\Delta_q + \tilde{\phi}^\Delta_q),
\end{eqnarray}
where $\phi^{SM}_q \equiv \arg{(-M^{q, SM}_{12}/\Gamma^{q, SM}_{12})}$,
$\phi^{SM}_s = 0.22^o \pm 0.06^o$, and $\phi^{SM}_d = -4.3^o \pm 1.4^o$ \cite{Lenz}.
Fits to decays $B^0_s (\bar{B}^0_s) \rightarrow J/\psi \phi$ and
$B^0_s (\bar{B}^0_s) \rightarrow J/\psi f_0$ determine, in particular,
$\Delta m_s$, $\Delta \Gamma_s$, and the angle \cite{Lenz}
\begin{equation}
-2 \beta_s + \delta^\textrm{peng, SM}_s + \delta^\textrm{peng, NP}_s
+ \phi^\Delta_s.
\end{equation}
If there is no CP-violation in the semileptonic decay of $B^0_q$ mesons, 
then $\delta^q_A = 0$. The angles $\delta^\textrm{peng, SM}_s$ 
and $\delta^\textrm{peng, NP}_s$ are due to penguin contributions
in the Standard Model and beyond.

In November of 2010, A. Lenz, U. Nierste and the CKMfitter Group \cite{CKMfitter}
presented detailed global fits to the Standard Model and to extensions of the
Standard Model (with $\delta^q_A = 0$, $\delta^\textrm{peng, SM}_s = 0$ and
$\delta^\textrm{peng, NP}_s = 0$). The global fit to the Standard Model finds
4 anomalies (i.e. parameters with ``pulls", defined in \cite{CKMfitter},
between 2 and 3 standard deviations):
$2 \beta_d$, $-2 \beta_s$, $A^b_{\textrm{sl}}$, and the branching ratio of
$B \rightarrow \tau \nu$. The pulls were 2.8, 2.3, 2.9 and 2.9
standard deviations respectively. The fits including the two independent
complex numbers $\Delta_d$ and $\Delta_s$ (with $\tilde{\Delta}_d = \tilde{\Delta}_s = 1$)
resolved all anomalies:
the pulls became 0.8, 0.5, 1.2 and 0.7 standard deviations respectively.
The fit obtained $|\Delta_d| = 0.747^{+0.195}_{-0.079}$, 
$|\Delta_s| = 0.887^{+0.143}_{-0.064}$, $\phi^\Delta_d = -12.9^{+3.8}_{-2.7}$ deg, and
$\phi^\Delta_s = -130^{+13}_{-12} \textrm{ or } -51.6^{+14.2}_{-9.7}$ deg.
It is noteworthy to mention that all 4 anomalies were corrected
by new physics in only the matrix elements $M^d_{12}$ and $M^s_{12}$. 

Since that review, there have been the following developments:
(i) the D\O\ Collaboration published the measurement with 9.0 fb$^{-1}$ \cite{D03};
(ii) the CDF and D\O\ Collaborations \cite{JCDF, JD} have presented new results on
$B^0_s \rightarrow J/\psi \phi$ in agreement with the Standard Model; and
(iii) the LHCb Collaboration has presented preliminary results, 
at the Lepton Photon Conference in September of 2011, on the decays
$B^0_s \rightarrow J/\psi \phi$ and $B^0_s \rightarrow J/\psi f_0$
with smaller uncertainties, and in good agreement with the Standard Model.

A detailed global fit including these new measurements is not yet available.
Let us here mention that these new measurements will add
to Figures \ref{1} and \ref{2}
a horizontal band that includes $a^s_{\textrm{sl}} = 0$, and are therefore not
in disagreement with the like-sign dimuon charge asymmetry, but shift
the burden of CP-violation towards the $B^0_d$ meson.

Finally, we should mention that new heavy particles, that could 
contribute to the Feynman box diagrams of $M^q_{12}$, have been
excluded up to higher energies by the Tevatron, and especially, by
the LHC (see summer conferences of 2011).

\section{\label{conclusion} Conclusions}
The D\O\ Collaboration has measured an anomalous
like-sign dimuon charge asymmetry that differs from
the Standard Model prediction by 3.9 standard deviations.
At present we do not understand the origin of this
discrepancy. The situation should become clearer
in the near future
when the following analysis become available:
(i) a new global fit to the Standard Model and beyond; 
(ii) the measurements of $A^b_\textrm{sl}$, $a^d_\textrm{sl}$ and
$a^s_\textrm{sl}$ with the final data sets of the Tevatron;
and (iii) the new results from the LHC
in both B-physics, and at the energy fronteer.
Or perhaps we will have new surprises?
We will work and see.

\bigskip 

\end{document}